\documentclass[acmsmall]{acmart}

\usepackage{booktabs} 

\acmDOI{10.1145/3342240}

\acmISBN{123-4567-24-567/08/06}

\newcommand{\cnn}{{\sc cnn}}
\newcommand{\fcl}{{\sc fcl}}

\usepackage{wrapfig}

\renewcommand\footnotetextcopyrightpermission[1]{} 
\begin{document}
\title{Take a Look Around: Using Street View and Satellite Images to Estimate House Prices
}

\author{Stephen Law}
\affiliation{%
  \institution{Alan Turing Institute and University College London}
  \streetaddress{96 Euston Rd}
  \postcode{NW1 2DB}
}
\email{slaw@turing.ac.uk}

\author{Brooks Paige}
\affiliation{%
  \institution{Alan Turing Institute and University of Cambridge}
  \streetaddress{96 Euston Rd}
  \postcode{NW1 2DB}
}
\email{bpaige@turing.ac.uk}

\author{Chris Russell}
\affiliation{%
  \institution{University of Surrey and Alan Turing Institute}
  \streetaddress{96 Euston Rd}
  \postcode{NW1 2DB}
}
\email{crussell@turing.ac.uk}


\begin{abstract}

  When an individual purchases a home, they simultaneously purchase its structural features, its accessibility to work, and the neighborhood amenities. Some amenities, such as air quality, are measurable whilst others, such as the prestige or the visual impression of a neighborhood, are difficult to quantify. Despite the well-known impacts intangible housing features have on house prices, limited attention has been given to systematically quantifying these difficult to measure amenities.
  Two issues have lead to this neglect. Not only do few quantitative methods exist that can measure the urban environment, but that the collection of such data is both costly and subjective.

  We show that street image and satellite image data can capture these urban qualities and improve the estimation of house prices. We propose a pipeline that uses a deep neural network model to automatically extract visual features from  images to estimate house prices in London, UK. We make use of  traditional housing features such as age, size and accessibility as well as visual features from Google Street View images and Bing aerial images in estimating the house price model. We find encouraging results where learning to characterize the urban quality of a neighborhood improves house price prediction, even when generalizing to previously unseen London boroughs.

 We explore the use of non-linear vs. linear methods to fuse these cues with conventional models of house pricing, and show how the {\em interpretability} of linear models allows us to directly extract proxy variables for visual desirability of neighborhoods that are both of interest in their own right, and could be used as inputs to other econometric methods. This is particularly valuable as once the network has been trained with the training data, it can be applied elsewhere, allowing us to generate vivid dense maps of the visual appeal of London streets. 

\end{abstract}

\keywords{real estate, deep learning, convolutional neural network, hedonic price models, computer vision, London}

\maketitle

\thispagestyle{empty}

\section{Introduction}
House pricing remains as much art as science. The cost of a property depends not just upon its tangible assets such as the size of the property and its number of bedrooms, but also on its intangible assets such as how safe or busy a neighborhood feels, or how a house stands with relation to its neighbors.
Real estate assessors face the challenging task of quantifying these effects and assigning to a property a realistic price that reflects what people are prepared to pay for these tangible and intangible assets.

From an economic perspective, it is unsurprising that people are prepared to pay for intangible assets. The urban environment directly effects people's social, economic and health outcomes. The design of a window placement can influence the amount of nature visible from within a home and also the perceived safety of a street~\cite{Jacobs1961}. The amount of greenery can influence both the pollutants at the street level and also its scenicness and ambiance~\cite{Seresinhe2017}. These differences in the urban environment are reflected in the varying prices people are prepared to pay in a property market, holding other factors such as size and access to jobs constant ~\cite{CheshireAndSheppherd1995,Law2016}.

\begin{figure*}[t]
  \includegraphics[width=1.0\linewidth] {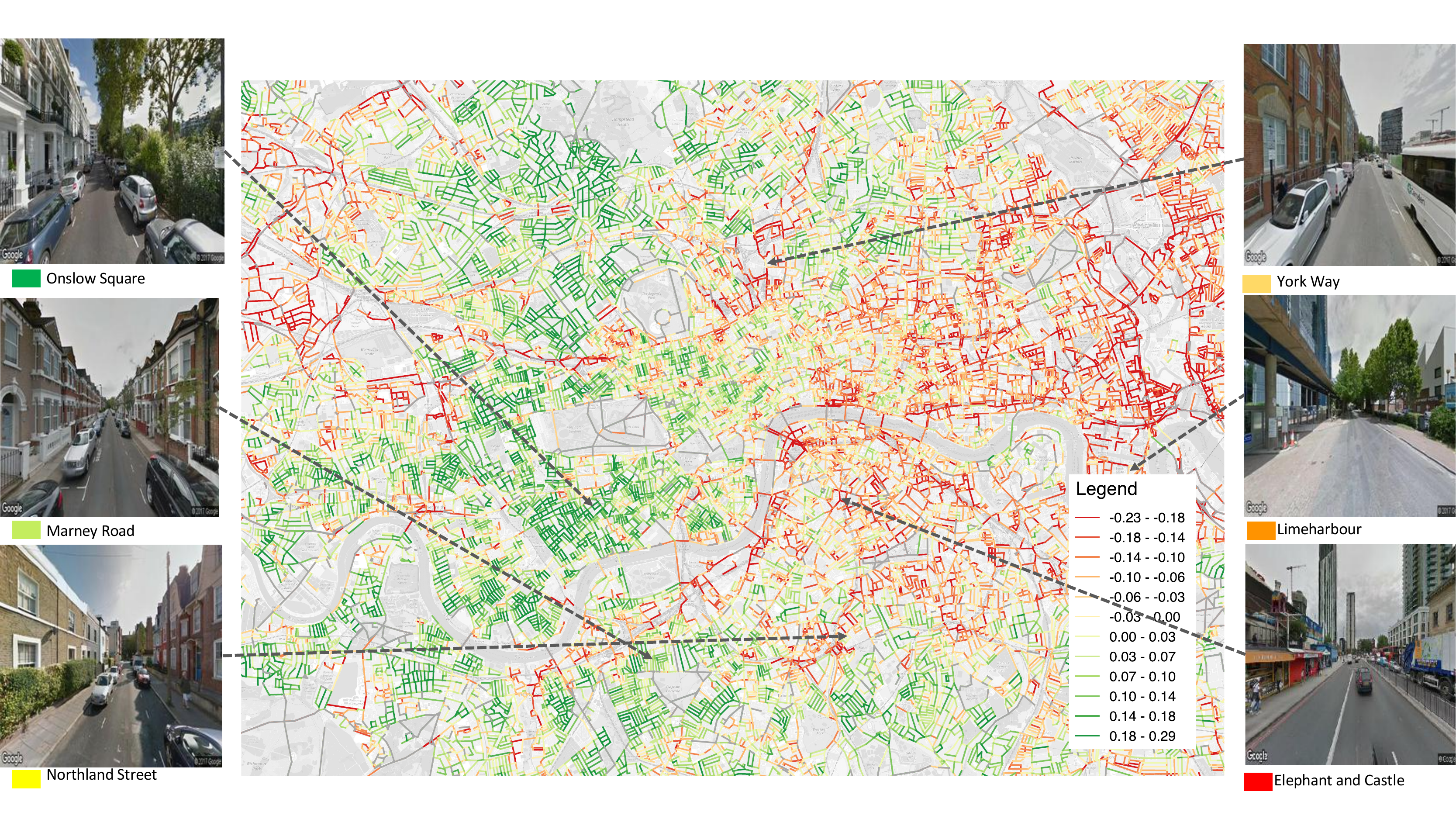}
  \caption{A map illustrating the latent visual appeal of neighborhoods across
    Greater London. Using a {\em linear hedonic model} we are able to extract
    the marginal effects of the visual appeal of the urban environment on house
    prices, as latent factors or proxy variables. The contribution of the urban
    environment retrieved from house prices varies from positive (green) through
    to negative (red). The map does not correspond to house prices,
    but to the visual appeal of the neighborhood which must be then combined with other housing attributes to price properties.  \label{fig:residue2}}
\end{figure*}

Some urban features are directly observable from photos, such as the activeness of a street frontage, the amount of greenery or the width of the pavement. Others are less directly quantifiable such as the prestige of the neighborhood or the aesthetics of the street. 
Despite the strong link between urban design attributes and economic value, there is a clear lack of research, computational tools, and data that can be used to discover these attributes and inform planning policies. To date, the discussion regarding which urban design attributes lead to better cities or higher property values has largely been theoretical, supported quantitatively by only a handful of studies. To measure these urban quality metrics require many street level surveys and structured interviews with professionals.

Collecting the data required to evaluate urban quality at the city scale is both costly and time-consuming.  One approach is to cast this as a problem of computer vision. This field has made great advances in image classification~\cite{Krizhevsky2012}, object detection~\cite{Girshick2015}, image segmentation~\cite{Chen2014} and edge detection~\cite{Li2016}. However, these advances have hinged upon the ready availability of {\em big data}, or in this context,  hundreds of thousands of diverse images annotated with these expensive quality metrics.

Unfortunately, this is a chicken and egg scenario: to avoid the expensive and time consuming hand annotation of images, we must first perform the expensive and time consuming process of hand-labeling of thousands of such images.  To avoid such issues, this research will not use machine vision methods to classify or to detect intermediate values, such as \emph{amount of greenery}, that can be used in house price models but instead use deep learning machine vision techniques to extract visual latent response in images based on the property price in an end-to-end learning model. We extract visual urban features using convolutional neural networks on urban images at both the plan and street-level which can be used in conjuncture with traditional housing features to estimate the price paid for a property in London.

A fundamental trade-off exists in econometrics between the use of tractable models~\cite{Rosen1974} that are easy to analyze, and difficult to interpret black-box approaches such as~\cite{You2017} that often have significantly better accuracy. To handle this dichotomy, we consider two different approaches; \emph{(i)} a full black-box model in which 
the 
neural network implicitly integrates the cues from standard attributes and from image data, and \emph{(ii)} a hybrid approach in which a mapping from the image space to latent attributes is handled by a \cnn\ and then the cues are fused by a standard linear model. This hybrid approach leads to the learning of interpretable semantic features that act as proxy variables for visual appeal of a neighborhood. Figure \ref{fig:residue2} shows a map of these features over greater London.

Our work differs markedly from previous research that has made use of images to price houses. First, we focus on using images of the urban environment at both the street and aerial level to estimate house prices rather than using interior images. More importantly, we have developed a set of interpretable proxy attributes which measure the visual desirability of neighborhoods; these variables can be used directly in existing econometric models. This concept is similar to the use of indices of multiple deprivation, crime attributes and school-performance data as proxy for neighborhood safety and prestige \cite{Gibbons2003}.

\section{Related work}
The cost of a heterogeneous good such as housing can be broken down into its utility-bearing components using the \emph{hedonic price approach}~\citep{CheshireAndSheppherd1995,Rosen1974}. The principle behind the hedonic price approach is that, holding all things constant, the influence of an attribute can be discovered by observing real estate values. One can imagine this concept by comparing two properties, each with nearly identical features, except that one property has one bedroom and the other has two bedrooms. The price differential between the two is equal to the implicit price of the extra bedroom. This approach can include structural features such as the size of a house, the age of a home, and the type of a home. It can also include location features such as employment accessibility or neighborhood features such as the number of shops nearby. Since its introduction, the hedonic price approach has become an established method for pricing environmental goods, constructing housing price indices, as evidence in the development of welfare policies and for usage in mass appraisal models ~\cite{Palmquist1984,Ridkerandhenning1967,McCluskey2012,McCluskey2013}. 

Despite the clear improvements in accuracy, there has been limited adoption of machine learning methods in house price estimation ~\cite{Peterson2009,Ahmed2016,You2017,McCluskey2013}. One reason is that the hedonic price approach can use the estimates of an ordinary least squares (OLS) model to recover the marginal willingness to pay for goods that are without explicit markets \cite{Rosen1974}. Despite its ease of use and interpretability, \citet{Peterson2009} argues such OLS model generate significant mispricing and misspecification errors; this motivates the use of a multi-layer perceptron (MLP) in a hedonic price model, which trades off interpretability for predictive accuracy\cite{McCluskey2013}.

In this paper we additionally adopt machine vision methods from street images to recover a visual latent response from the urban environment. \footnote{
Note that we define visual latent response here as an overall response of the visual system that is semantically interpretable. This notion differs to visual feature which is commonly defined in the machine vision community as an important patch of the image that drives a classification response.}

Recent studies from \citet{Naik2013}, \citet{Liu2017}, and \citet{Law2017} have began to leverage on the availability of large scale street image data to extract urban knowledge.  For example, both \citet{Liu2017} and \citet{Law2017} used machine vision techniques to retrieve geographical knowledge such as street frontage quality.  In contrast, \citet{Naik2013} used Street View images to estimate the perceived safety of streets \cite{Streetscore}. This research is related to this latter study in extracting a global statistic from street images.

\begin{figure}
	\centering 
	\includegraphics[width=0.9\linewidth]{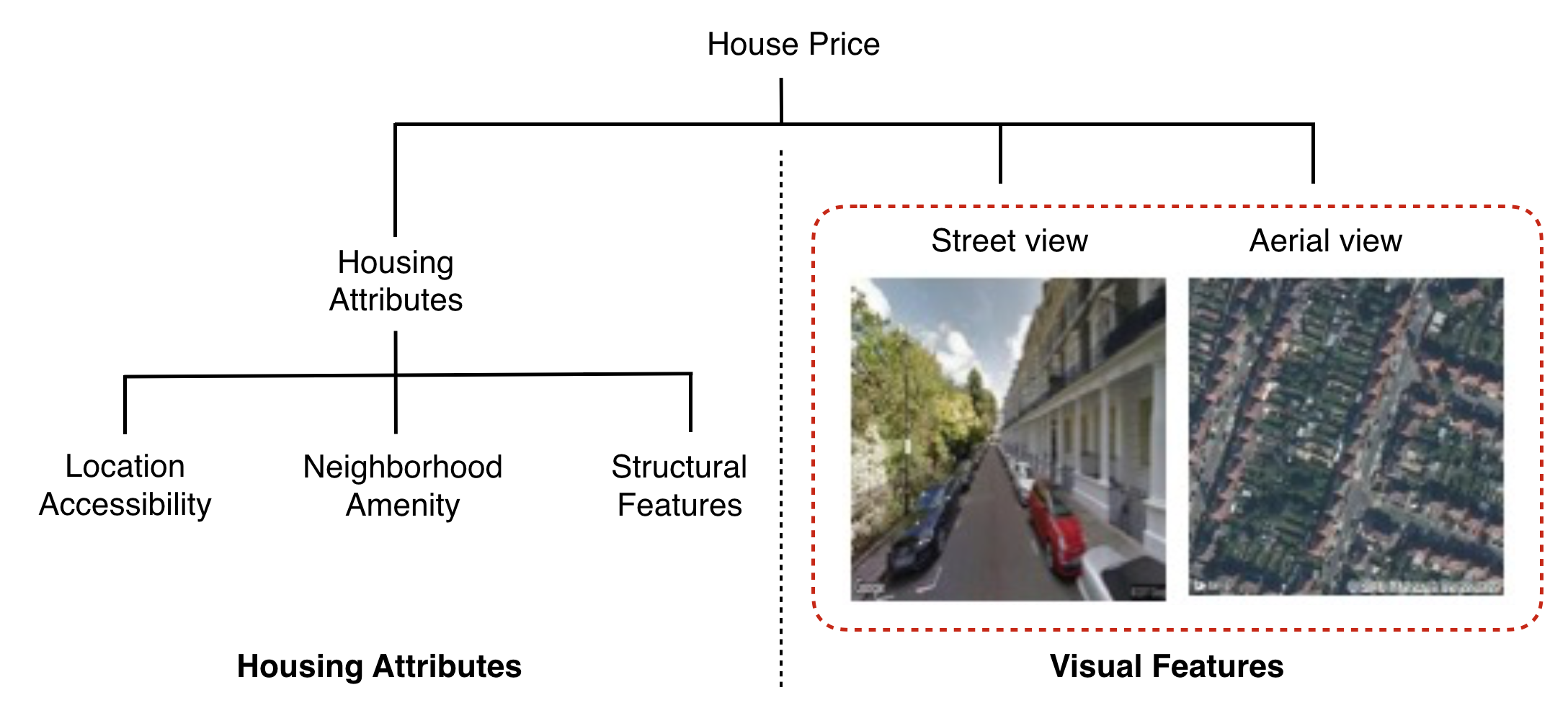} 
	\caption{Conceptual model showing how urban visual latent response can be integrated
    with a traditional house pricing model.}
	\label{fig:external}
\end{figure}

The following section will outline the current status of  machine learning techniques in house price estimation.
The first is a study from \citet{Peterson2009} that used a multi-layer perceptron model to estimate house price with traditional housing features such as age, size, accessibility and safety. The author compared an artificial neural network hedonic price model with two hidden layers to a standard OLS hedonic price model.  The author found significant improvements in the use of an ANN model. The improvements are unsurprising due to the expected nonlinear relations captured by variables in the hidden layers which cannot be modeled by a linear OLS model. 

These nonlinear effects become more important when dealing with intangibles such as the quality of the neighborhood, as these intangibles can often have a multiplicative effect on the hedonic value assigned to tangible assets. For example, each square meter of property could cost orders of magnitude more in an exclusive neighborhood than in a run down one. 

A study from \citet{Ahmed2016} supplemented traditional housing features with image features extracted from property photos. The study used both property photos and traditional housing features in estimating house price. The result found objects identified using traditional machine vision methods such as Speeded Up Robust Features (SURF)~\cite{Bay06surf:speeded} significantly improved the model. The research also compared a support vector regression model to a neural network model and found that the neural network one achieved better results.

\citet{You2017}  also used image features extracted from property photos to estimate a house price model. Instead of using traditional machine vision techniques in detecting image features, this research made use of a novel recurrent neural network LSTM model to predict house price using the image features from its wider district neighbours (second order spatial effect) in an end-to-end learning model. 

\begin{table}[]
\centering
\caption{
Descriptive statistics}
\label{table:descriptive}
\begin{tabular}{rrrrrr}
\toprule
{} &    n& mean &   sd &   min &   max \\
\hline
log price &   40,470&12.03 &  0.62 &  0.69 &  15.3 \\
year    &     40,470&0.42 &  0.22 &  0.00 &   1.0 \\
size    &     40,470&0.52 &  0.14 &  0.00 &   1.0 \\
beds    &     40,470&0.30 &  0.14 &  0.00 &   1.0 \\
age     &     40,470&0.62 &  0.14 &  0.00 &   1.0 \\
type    &     40,470&0.30 &  0.41 &  0.00 &   1.0 \\
park    &     40,470&0.76 &  0.15 &  0.00 &   1.0 \\
shops   &     40,470&0.46 &  0.19 &  0.00 &   1.0 \\
gravity &     40,470&0.65 &  0.14 &  0.00 &   1.0 \\
\hline
\hline
street photo & 111,701 \\
aerial photo & 111,701 \\
\hline
\end{tabular}
\end{table}

\citet{Gebru2017}  extracted car types, years, and make from 50 million Google Street View images to correlate with socio-economic factors such as income and geographic demographic types across different cities in the United States. The study found that car types, years, and makes can be used as features to predict accurately the income, race, education, and voting patterns at both the zip code and precinct level. 

Several related works do not model house prices directly, but provide further evidence that street-level photographs of a city can be used to estimate relevant features. 
\citet{Dubey2016} collected human perception data from street images (Place Pulse 2.0) through a crowd-sourced survey \cite{Naik2013}. They then fit a model to predict these human perception factors, such as perceived safety and liveliness, directly from the images; these factors are likely important covariates in a house price model.

\citet{Arietta2014} presented a method for automatically identifying and validating predictive relationships between the visual appearance of locations in a city and properties such as theft rates, house price, population density, tree presence, and graffiti presence. The novelty of the study is it extracted a set of discriminative visual features \cite{Doersch2012} such as roof types and window types that corresponds to a location attribute using a support vector machine. The model successfully identified visual features that corresponds to location with higher or lower house price (binary). However the model did not generalize well across cities in the States.

We differ from these previous approaches in multiple ways. First, this study collects urban neighborhood images~\cite{Gebru2017} both at the street level and aerial level rather than images of the property itself ~\cite{Ahmed2016,You2017,Poursaeed2018}. This research allows the neighborhood features to be extracted from two perspectives, the street of the property and the neighborhood surrounding the property (Figure~\ref{fig:external}). Secondly, we compare a ANN hedonic price model~\cite{Peterson2009} with only housing features to a model that is augmented with both street images collected from Google
Street View API~\cite{GoogleStreetView}, and aerial images collected from Bing Images API. Third, we compare a nonlinear hedonic price model with a hybrid linear hedonic price model where the images gets encoded into a latent variable that achieves greater interpretability. Finally, the model is tested
on multiple neural network architectures and using a spatial out-of-sample
testing set, in which Southwark, an entire borough of London, was excluded from the training set, to demonstrate the generalizability of the results.

\section{Method and Materials}

\begin{figure*}
	\centering
	\begin{tabular}{p{0.49\textwidth}p{0.49\textwidth}}
	  \includegraphics[width=0.49\textwidth] {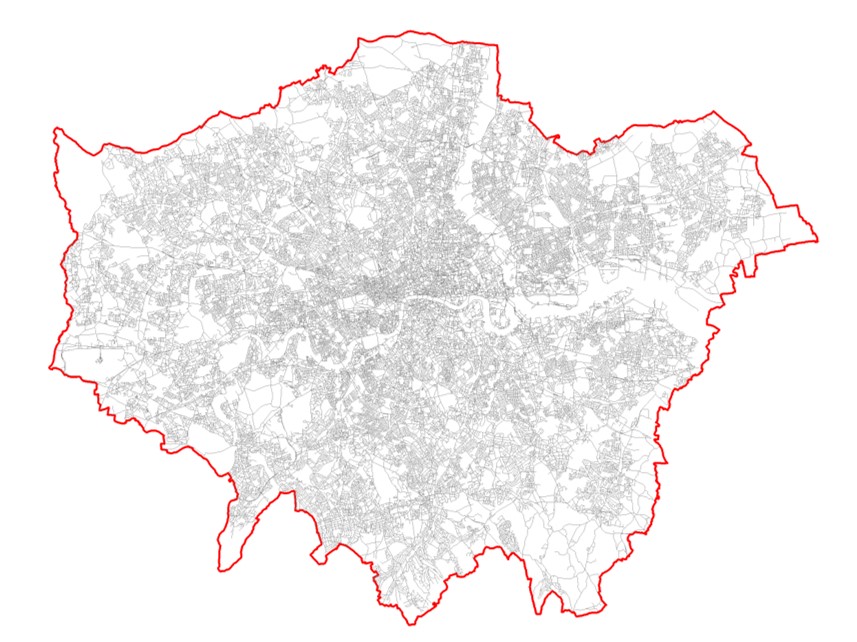}
    &
	  \includegraphics[width=0.49\textwidth] {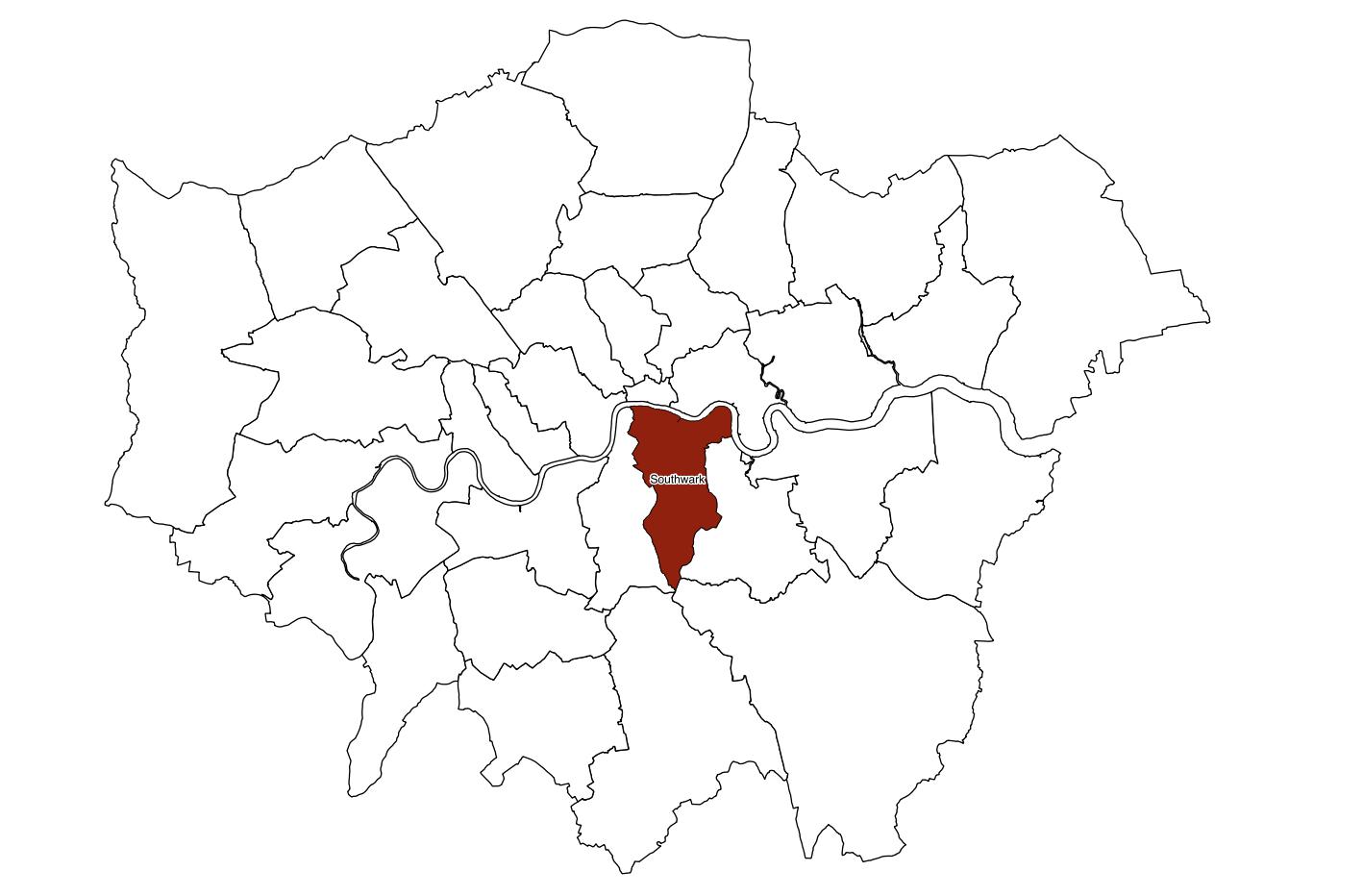}
  \end{tabular}
	\caption{{\bf Left:} Greater London study area. {\bf Right:} The Southwark test-set used in one of the experiment.
	\label{GLA}	\label{fig:southwark}
 Contains Ordinance Survey data \copyright Crown copyright and database right \copyright 2017.}

\end{figure*}

We propose a model which estimates the log house price from three separate sets of input data: housing attributes, street images, and aerial images.
To demonstrate the effectiveness and utility of our model, we use Greater London (Figure~\ref{GLA}) as a case study. 
The procedure consists of a data collection phase, a training phase and a testing phase; we begin by describing the data collection phase.

\subsection{Data collection}
We collected three datasets in the data collection phase. The first dataset is comprised of traditional housing attributes including structural, neighborhood and location features. 
House price data is taken from the UK Land Registry Price Paid dataset \cite{LandRegistry}, which includes transaction details for all property sales in England, with additional property attributes from the Nationwide Housing Society \cite{Nationwide}.
The structural features, for each property transaction, include the location of the property, the price paid for the property, the type of the property, the size of the property, and the age of the property. Location features include gravitational accessibility to employment. The statistic was computed as a gravity model, where accessibility is measured as a sum of jobs divided by distance within 60 minutes, $\sum{e_{ij}}d_{ij}^{-1}$.

\begin{wrapfigure}{r}{0.5\textwidth} 
	\centering 
	\includegraphics[width=0.7\linewidth] {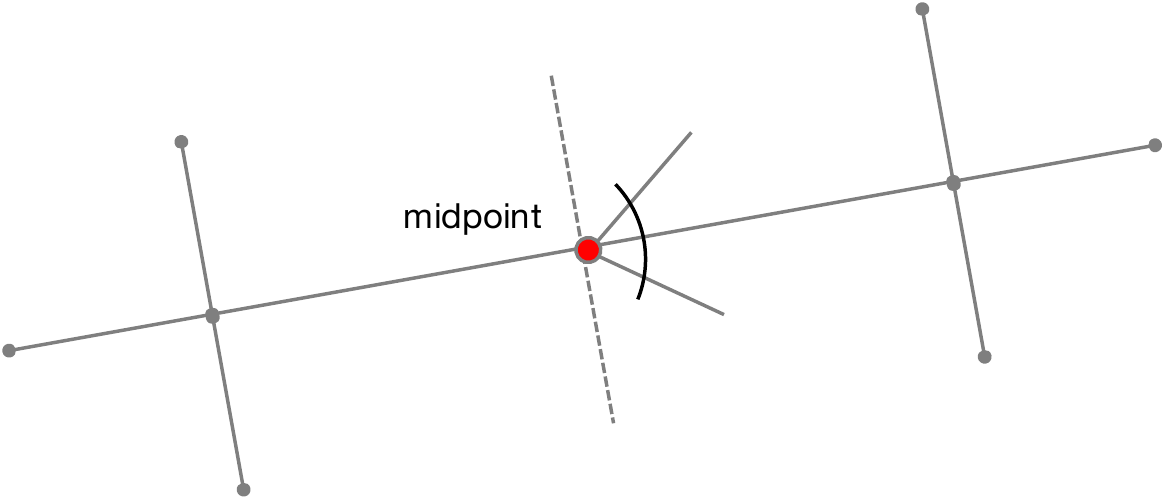}
	\caption{Street bearing diagram, and the orientation of front facing cameras.}
	\label{fig:Bearing}
\end{wrapfigure}

Neighborhood features include the distance to the nearest parks and the number of shops and commercial uses within 800 meters. The datasets used to calculate these location features come from the Ordinance Survey \cite{OS}, the Office for National Statistics \cite{ONS} and Historic England \cite{HistoricEngland} . This dataset consists of a total of $130,557$ transactions which are then grouped to the nearest street. This procedure results in 40,470 streets having at least one property transaction out of a total of $110,701$ streets. Descriptive statistics are shown in Table~\ref{table:descriptive}. The output variable, price, is log transformed, while all the input attributes are log transformed and then linearly rescaled to have minimal values of $0$ and maximal values of $1$. 

\begin{figure}[t]
\centering
\begin{minipage}[t]{0.44\linewidth}
\vspace{0pt}
	\caption[justification=raggedright]{Data sources for the visual latent response. \\
    (Top row) Valid Google Street View images. \\
    
    (Middle row) Invalid images discovered using techniques in \cite{Law2017}. From left to right: not available image; dark image; interior image; interior image.
    \copyright 2017 Google Inc. Google and the Google logo are registered trademarks of Google Inc. \\
    
    (Bottom row) Microsoft Bing aerial images, \copyright 2018 Microsoft.
    \label{fig:frontages}	\label{fig:invalid}	\label{fig:aerial}
    }
\end{minipage}\hfill
\begin{minipage}[t]{0.52\linewidth}
\vspace{0pt}
  \begin{tabular}{p{\linewidth}}
	\includegraphics[width=1\linewidth] {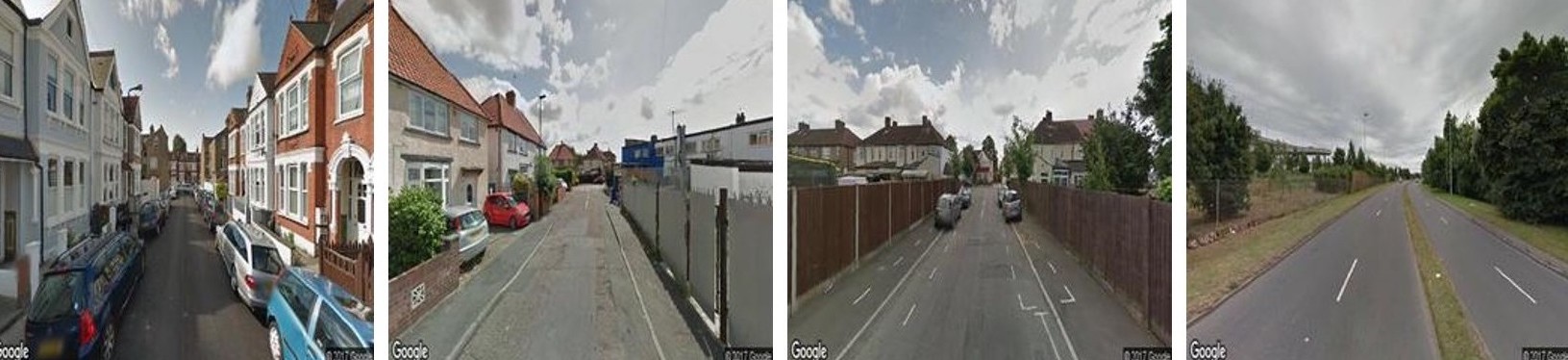}\\
  \includegraphics[width=1\linewidth] {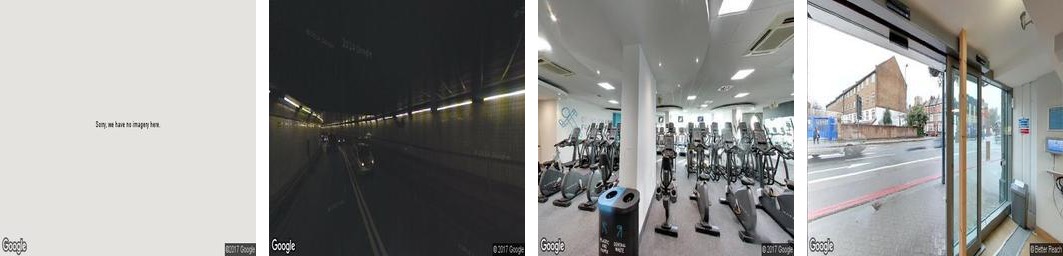}\\
	\includegraphics[width=1\linewidth] {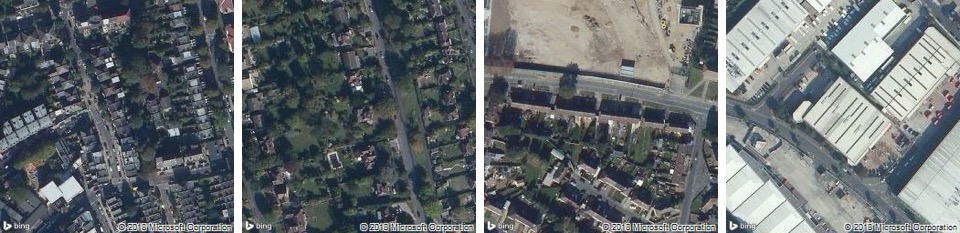}
  \end{tabular}
\end{minipage}
\end{figure}

The second dataset is comprised of street images taken from the Google Street
View API~\cite{GoogleStreetView}\footnote{\copyright 2017 Google Inc. Google and the Google logo are registered trademarks of Google Inc.}. Following \cite{Law2017}, one front-facing image was collected for each street in the Greater London Area using the API. (A front facing image is one which faces towards the front of the car, i.e.\ it typically faces away from the property at
a ninety degrees angle; see Figure \ref{fig:Bearing} for clarification.) To
collect the dataset, we first constructed a graph from the street network of
London (OS Meridian line2 dataset \cite{OS}), in which every node is a junction
and every edge is a street. We then take the geographic median and the azimuth
of the street edge to give both the location and the bearing when collecting
each image. This process is to ensure the Street View images are
constantly front-facing and are taken from the center of the road. This reduces
the problem of images being too close to the junction. The field of view has been set to 120 degree in order to ensure that both sides of the building facades are captured. A typical data cleaning procedures is then undertaken which includes removal of invalid images such as the interior of buildings, images that are perpendicular to the street and not front-facing, images that were too dark or those not
available, using a series of automatic functions, CNN classifiers and manual processes~\cite{Law2017}. Figure \ref{fig:frontages} (top two rows) shows examples of the valid and invalid Street View images. 
A limitation is this process does not capture all invalid images such as those blocked by large vehicles.

The third dataset is comprised of aerial images extracted from the Microsoft Bing Images API~\cite{BingMaps}\footnote{\copyright  Bing. All rights reserved.}. Using the API, one aerial image has been collected for each street in the Greater London Area. To collect the dataset, we take the centroid of each street edge from the OS Meridian line 2 dataset \cite{OS}. We then download for each street an aerial image with a zoom level parameter set at 18 (roughly 150m) to get a constant aerial view of the street neighborhood. Figure \ref{fig:aerial} (bottom row) shows examples of these aerial images.

We collected a total of 111,701 images for both the street image dataset and the aerial image dataset. 40,470 of the images have at least one property transaction. Both sets of images were re-sized into a uniform dimension (256 pixels x 256 pixels). 

\subsection{Visual Feature Extraction}

\begin{figure*}[t]
\centering 
\begin{minipage}[b][][b]{.49\textwidth}
    \centering 
	\includegraphics[width=0.95\textwidth] {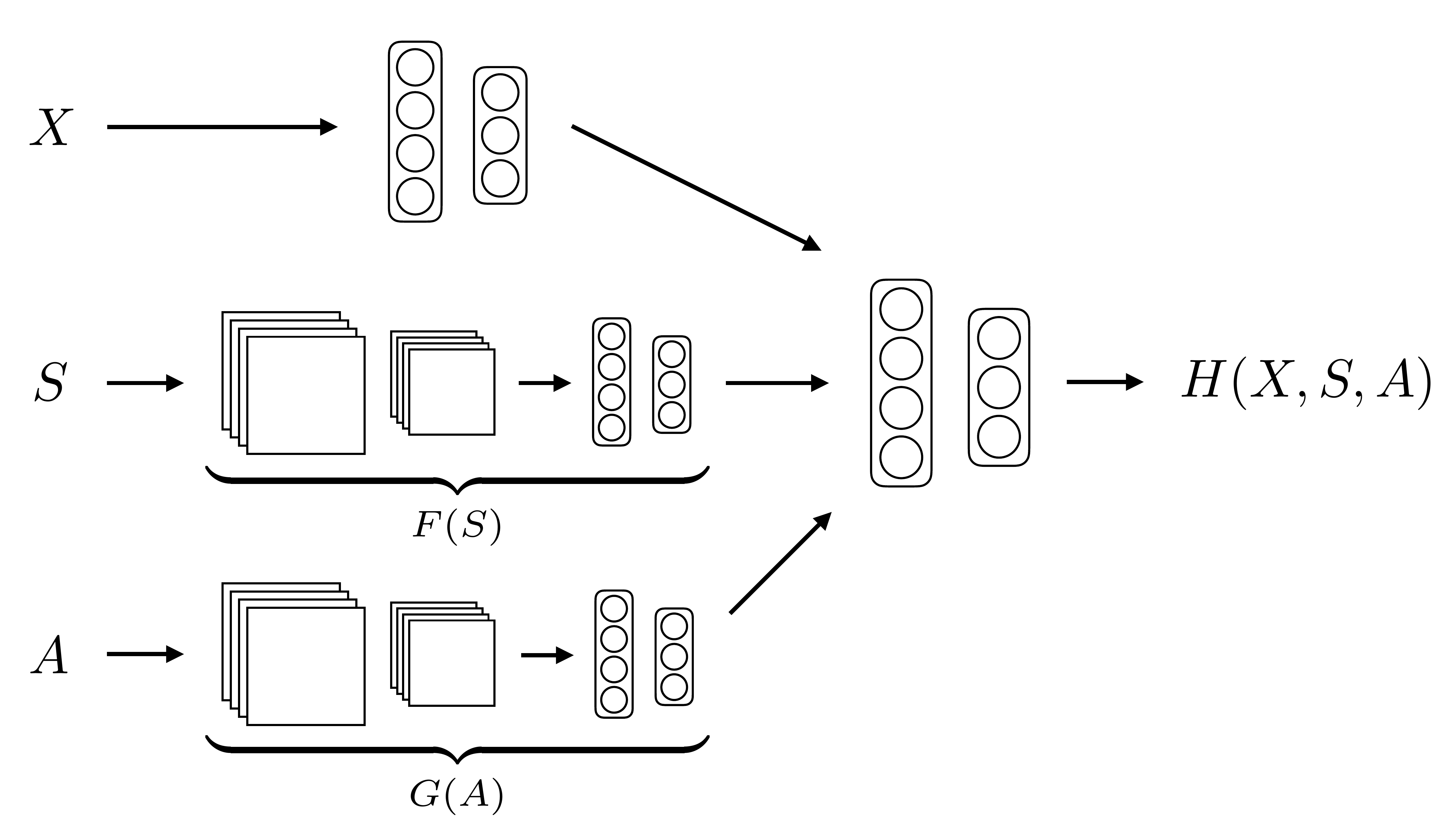}
	\captionof{figure}{Fully nonlinear model network structure.}
    \label{fig:network_struc}
\end{minipage}\hfill
\begin{minipage}[b][][b]{.49\textwidth}
	\centering 
	\includegraphics[width=0.95\textwidth] {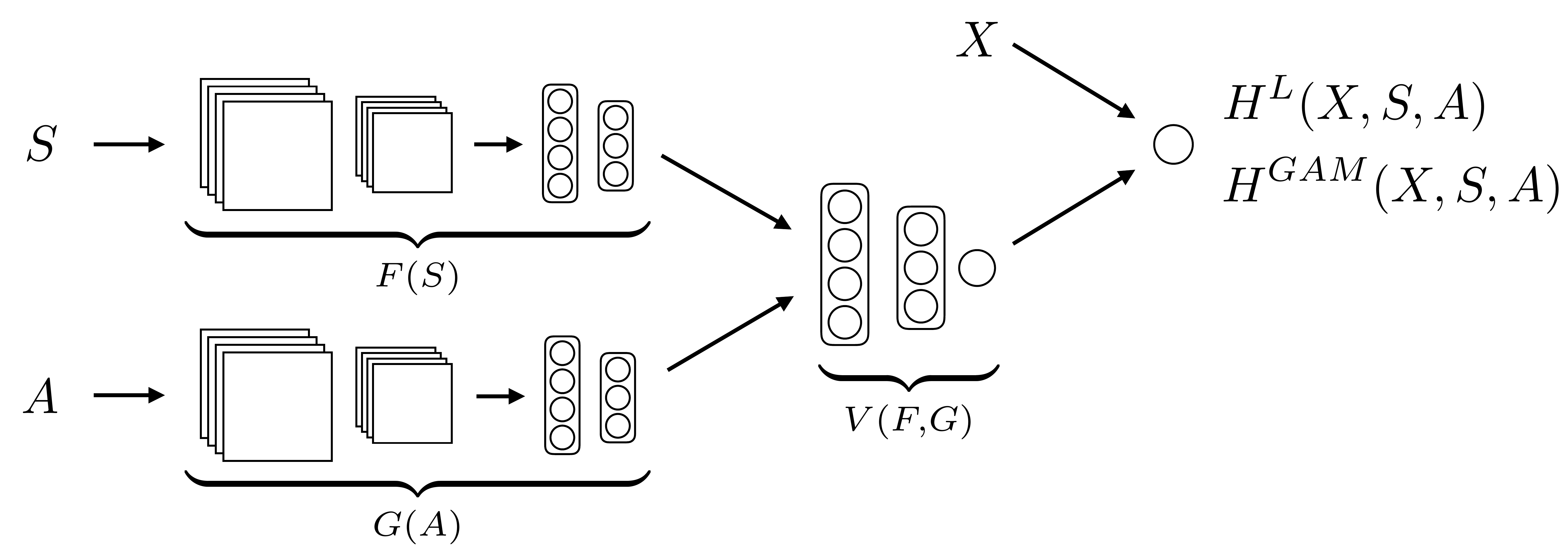}
    \vspace{1.5em}
	\captionof{figure}{Semi-interpretable model network structure.}
    \label{fig:network_struc_linear}
\end{minipage}
\end{figure*}

We depart from both standard OLS models and the hedonic perceptron model of~\citet{Peterson2009} in that we also allow the input of a latent attribute that can be understood as a proxy for the desirability of the urban environment, as captured by Street View data and by satellite imaging.

This proxy variable is given by the responses of an additional neural network. Importantly, as we do not have expert annotations of the desirability of the urban environment, we learn feature extractors for the Street View and satellite imagery by composing these networks with a hedonic price model $H(\cdot)$ and training the entire architecture end-to-end, while controlling for the contribution of the individual housing attributes.

To extract meaningful features from the Street View images $S$ and aerial photos $A$, we define two functions $F(W_S, S)$ and $G(W_A, A)$, parameterized respectively by weights $W_S$ and $W_A$; for notational simplicity we will generally refer to these simply by $F(S)$ and $G(A)$.
Although the functions have different weights, both networks adopt the same convolutional neural network (\cnn) architecture for the vision model.
In a \cnn\ model, the earlier layers detect the basic edges while the ladder layers detect the more complex shapes. The model follows the basic \cnn\ architecture that uses 3x3 filters that are tested on $4$, $8$, and $13$ convolutional layers follow by a series of pooling layers (as in e.g. VGG\cite{VGGnet}). 
We take the value at the final flattened convolutional layer as the output of the \cnn.
These outputs are feature vectors which summarize the Street View and aerial photo data, respectively, and can be used alongside the data $X$ as additional inputs into a hedonic price model.

\subsection{Model Architectures}

There are two important uses of such hedonic models. The first lies in accurately predicting house prices as a guide for realtors and for people looking to put their house on the market; for such individuals, accurate pricing is the most important criteria, and they are happy with the use of black-box models such as neural networks providing they lead to improve accuracy. The second use of these models lies in econometrics; here interpretability and ease of analysis are more important than accuracy and the use of linear model is still favored.

Because of this, we consider two general forms of the hedonic price model.
The first form is designed to maximize the predictive accuracy of $H(\cdot)$, which is a multi-layer perceptron (see Figure~\ref{fig:network_struc}), capable of learning arbitrary functions. This model can capture any possible interactions between the variables and nonlinearity in the responses.
We also consider alternative models which restricted $H(\cdot)$ to only linear or additive functions of $X$ (see Figure~\ref{fig:network_struc_linear}); the linear model permits only a weighted combination of features, while the additive model is slightly more general in that it learns nonlinearities, but not interactions. 
In both cases, the sub-networks $F(S)$ and $G(A)$ are multi-layer convolutional networks capable of learning nonlinear responses which are used to process the Street View and aerial images.

Our objective in these experiments is to understand the effect of visual latent response on house price: to what extent is the price of a house explained by the visual appeal of the immediate area? We test this using both interpretable models (linear and additive) which are designed for explainability, as well as a hedonic perceptron model \citep{Peterson2009} which is designed for maximum prediction quality. These models are considered both with and without the visual latent response, which allows us to measure the improvement in predictive accuracy in different settings. These linear, additive, and perceptron models form a set of baselines for our experiments, where we examine how well an extracted visual desirability can improve predictive performance. \color{black} 

\subsubsection{The Hedonic Perceptron Model} 
The fully nonlinear model depicted in Figure~\ref{fig:network_struc} can be understood as a natural generalization of the hedonic perceptron model used by works such as~\cite{Peterson2009}. 
We train a multi-layer neural network to predict log house prices on the basis of a set of normalized attributes (see Table~\ref{table:descriptive}). 

We represent the overall price of the property by a function $H(W_H,\cdot)$, parameterized by a set of weights $W_H$ that takes as input housing attributes $X$ and extracted image features $F(S)$ from Street View images $S$ and $G(A)$ from aerial photos $A$.
For purposes of establishing baselines and quantifying the relative predictive capability of the housing attributes and the new image data, we also perform an ablation study in which we consider a baseline model $H(W_H, X)$ that depends only on the housing attributes $X$, as well as models $H(W_H, \dots)$ 
which can additionally incorporate either or both of the Street View and aerial photos; the full combination of experimental setups is described in Section~\ref{sec:experiments}.

For the nonlinear hedonic perceptron model, a fully connected neural network with two hidden layers is adopted.
The first fully connected layer (\fcl) has 128 hidden nodes, while the second \fcl\ has 64 hidden nodes. 
This layer represents an extracted feature vector with a nonlinear dependence on $X$.
In the baseline model $H(W_H, X)$, a final \fcl\ outputs the overall response of the model; 
for the models which include the images $S$ and/or $A$, we concatenate this vector to vector-valued output of the functions $F(\cdot), G(\cdot)$ and use this as input into an additional fully-connected network, again with two hidden layers of 128 and 64 nodes respectively. 
These taken together yield an overall combined nonlinear predictive model of the form
\begin{equation}\label{eq:merge1}
H(X,S,A)=H(W_H,X,F(W_S, S),G(W_A, A)).
\end{equation}
The difference between the predicted log price $\hat Y = H(X,S,A)$ given by Equation~\eqref{eq:merge1} and the actual log price $Y$ is given by the mean squared error loss function
\begin{equation}\label{eq:loss1}
L(W_H,W_S,W_A) = \frac{1}{n}\sum\big(Y-H(X,S,A)\big)^2.
\end{equation}
This loss is a function of the weights $W_H,W_S,W_A$ which are optimized in the learning process, via ADAM which is an adapted stochastic gradient descent optimiser with a learning rate of 0.001 for 80 epochs \cite{ADAM2014}. 
\color{black}

\subsubsection{Linear Hedonic Model}

The linear hedonic model can be interpreted as a network with no hidden layers, that consists of a single neuron with no nonlinearity which directly outputs the response.
A primary difference between the linear and nonlinear models is in the handling of the images themselves. In the nonlinear model, the trained sub-networks $F(\cdot)$ and $G(\cdot)$ extract a feature {\em vector} when used as inputs to the nonlinear model. 
In the linear model, we insert an additional network $V(F, G)$, defined by two additional fully-connected layers with weights $W_V$, which compress the feature vectors output by the \cnn{s} to a single scalar response.
This scalar summary can then be included as additional independent variable in an OLS model, where it functions as a proxy variable to control for visual desirability of the local urban environment.
We can then compare a standard linear model
\begin{align}\label{eq:hedonic1}
H^L(X) &=\beta_0 + \sum{\beta X}+\epsilon,
\end{align}
which only uses the housing attributes $X$,
with an extended model $H^L(X,S,A)$ that includes the visual desirability as
\begin{align}
%
H^L(X,S,A) &=\beta_0 + \sum{\beta X}+\gamma V(F(S), G(A)) + \epsilon,
\label{eq:hedonic2}
\end{align}
where $\beta$ are the OLS regression weights, and the extended model includes an additional weight $\gamma$.

One benefit of the interpretable econometric approach is that the learned feature response $V(F, G)$ can be directly interpreted as a measure of how the visual desirability of the neighborhood alters the value of the house prices. Figure~\ref{fig:residue2} shows a heat plot of these responses over the whole of London.

\subsubsection{Generalized Additive Model} 

As a middle ground between the simple linear model and the general hedonic perceptron model, we also consider including these visual scalars in more generalized regression models. 
We investigate using these visual scalar outputs as inputs into a generalized additive model \cite{Wood2006},
in which each of the individual housing attributes $x_d$ is transformed by a univariate nonlinear function $f_d(x_d)$, typically taken to be defined by a spline basis.
This yields a model of the form
\begin{equation}\label{eq:GAM}
H^{GAM}(X, S, A) = \beta_0 + f_1(x_1) + f_2(x_2) +... + f_D(x_D) + f_v(V(F(S), G(A))) + \epsilon,
\end{equation}
where each of the individual housing attributes, as well as the visual desirability proxy variable, are subject to a potentially nonlinear transformation.
This model includes the linear regression model as a special case, in which each $f_d(x_d)$ is a constant multiplicative function.

This family of model represents a natural tradeoff between the interpretability of the OLS model and the predictive accuracy of black-box machine learning models \citep{lou2012intelligible}.
We use the same learned proxy variable as used above in the linear model; the purpose of the additive model is to explore how far we can push predictive accuracy while maintaining model intelligibility.

\section{Experimental Results}
\label{sec:experiments}
We consider four sets of experiments: the first two using general neural networks to regress, and the last two using a standard OLS linear regressor or generalized additive model, with neural networks as mid-level components.

\subsection{Spatially Missing-at-Random}

The first two sets of experiments are designed to test the predictive performance of the hedonic perceptron model, and particularly to quantify the importance and contribution of the street view photos and aerial imagery.

To test the importance of particular attributes with respect to the model accuracy, we constructed six different models. The first three models are individual models for each data source. The final three models are different combinations of multiple data sources.
For all experiments, we train the model end-to-end to minimize the mean squared error on a training set, using the ADAM optimizer with the default initial learning rate set at 0.001. We report two test set metrics: the mean squared error (MSE) and the coefficient of determination $R^2$ between the model prediction and the actual log-price. All the experiments are conducted with the Keras library~\citep{chollet2015} using a Tensorflow~\cite{tensorflow2015-whitepaper} back-end.

In the first experiment, we tested three variations of these six models by altering the architecture of the Street View network $F(S)$ and the aerial imaging network $G(A)$. We split the dataset randomly where 70\% is used for training, 15\% is used for validation, and 15\% is used for testing, yielding an experimental setting in which the test set is spatially missing-at-random relative to the training set.
We tested a 4-layer \cnn, a 8-layer \cnn\ and a 13-layer \cnn\ model.
Note that varying the architecture does not alter the attribute-only model, which has no convolutional layers.

\begin{figure}
\centering
\includegraphics[width=0.49\linewidth]{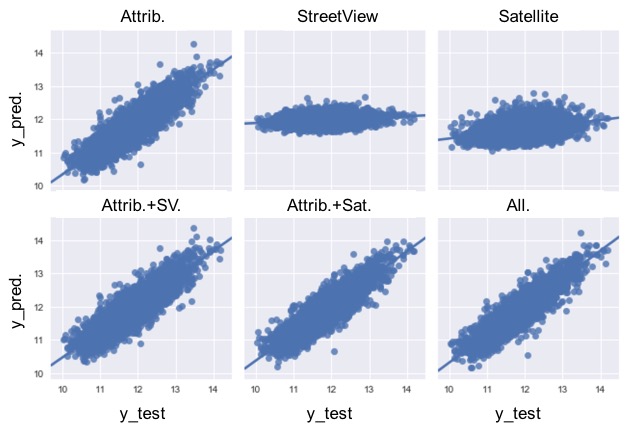}
\hfill
\includegraphics[width=0.49\linewidth]{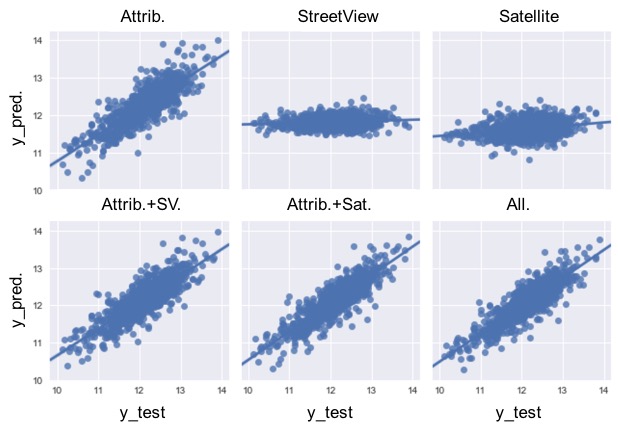}

\caption{{\bf Left:} Scatter-plots showing correlations for each model on the spatially missing-at-random experiment. {\bf Right:} Scatter-plots showing the correlation for each model on the holding-out Southwark experiment. 
\label{fig:scatterplot}\label{fig:southwark_scatterplot}\label{my-label}}
\end{figure}

Figure \ref{fig:scatterplot} shows the scatter-plots between the actual and the predicted log price for all six models, using the best-performing architecture. The result shows quite clearly that the four models which include the housing attributes $X$ as one of the inputs achieve much higher correlation than the two models which use only Street View or aerial image data. This is to be expected, as these models only have visual information for the prediction model.

\begin{figure*}
\includegraphics [width=0.49\textwidth]{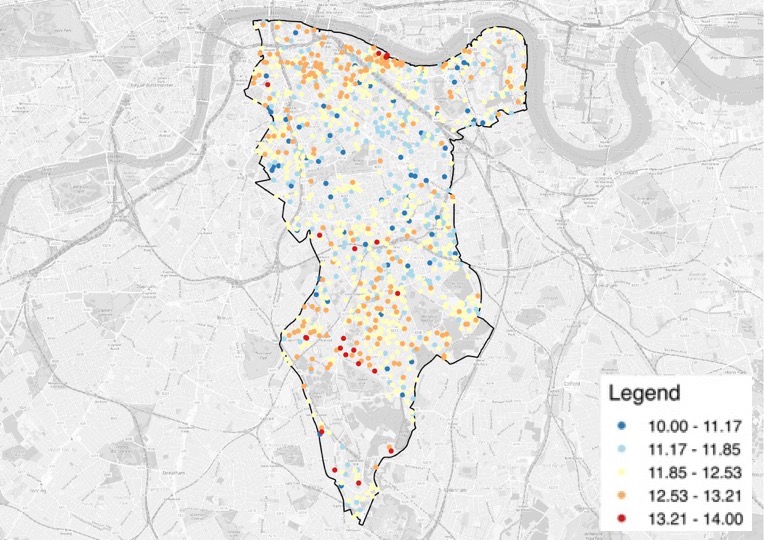}
\hfill
\includegraphics [width=0.49\textwidth]{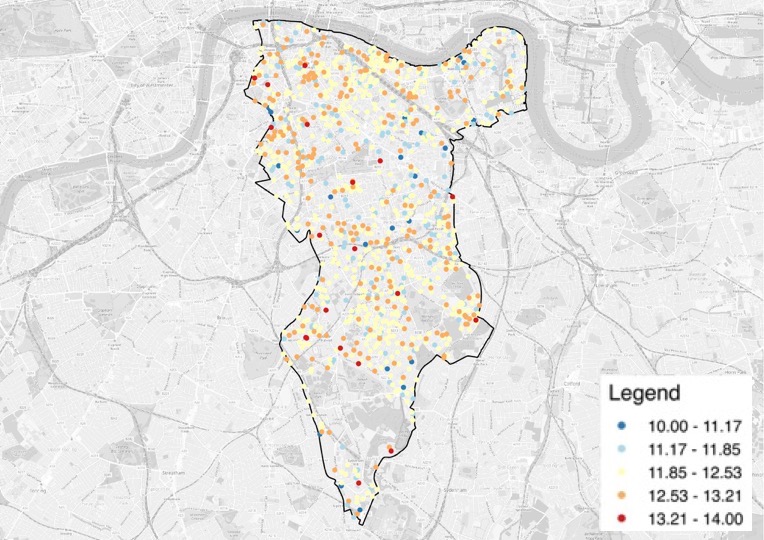}
  \caption{Test of generalization ability: predicting prices in the Borough of Southwark, using a model trained on data from elsewhere in London. {\em Left:} Actual log-price; {\em Right:} Predicted log-price.
  Survey data \copyright Crown copyright and database right \copyright 2017.\label{fig:ssn_gla}}
\end{figure*}

The result in Table~\ref{tab:run1} shows the mean squared error and $R^2$ for all six models, and across all three sizes of architectures. Of the single data source models, the housing attribute model achieve better accuracy than both the Street View model and the aerial-image model.  Models using multiple data sources achieves better accuracy than the single data source models. The model with both $X$ and $S$ achieves 83\% accuracy, while the model with both $X$ and $A$ achieves 84\% accuracy and the full model, including all of $X$, $S$, and $A$ achieves 85\% accuracy.  The results show that the model that combines housing attributes with the images achieves a better result than the one without. 

\begin{table}[]
\begin{minipage}[b][][b]{.49\textwidth}
\centering
\caption{Spatially missing-at-random results.\\
  MSE (top) and $R^2$ accuracy (bottom)
  \label{tab:losses}\label{tab:run1}}
\resizebox{.95\textwidth}{!}{%
\begin{tabular}{r|c|c|c}
       MSE         & 4-layers & 8-layers & 13-layers \\ \hline
Attributes only    & 0.08 & -- & -- \\ \hline
Street View only   & 0.33 & 0.32 & 0.31 \\ \hline
Aerial only       & 0.34 & 0.29 & 0.31 \\ \hline
Attrib.\ + Street   & 0.06 & 0.06 & 0.06 \\ \hline
Attrib.\ + Aerial   & 0.06 & 0.06 & 0.05 \\ \hline
Full model          & 0.06 & 0.05 & 0.05 \\ 
\end{tabular}}
\vspace{1em}
\\
\resizebox{.95\textwidth}{!}{%
\begin{tabular}{r|c|c|c}
       $R^2$     & 4-layers & 8-layers & 13-layers \\ \hline
Attributes only  & 80.74 & -- & -- \\ \hline
Street View only & 4.92 & 5.16 & 9.59 \\ \hline
Aerial only     & 15.22 & 15.75 & 15.95 \\ \hline
Attrib.\ + Street & 81.68 & 81.34 & 82.97 \\ \hline
Attrib.\ + Aerial & 81.91 & 83.58 & 84.45 \\ \hline
Full model        & 83.10 & 83.69 & 84.67 \\ 
\end{tabular}}
\end{minipage}\hfill
\begin{minipage}[b][][b]{.49\textwidth}

\centering
\caption{Generalization to held-out Southwark.\\ 
MSE (top) and  $R^2$ accuracy (bottom)
\label{tab:run2}}
\resizebox{.95\textwidth}{!}{%
\begin{tabular}{r|c|c|c}
       MSE         & 4-layers & 8-layers & 13-layers \\ \hline
Attributes only     & 0.13 & -- & -- \\ \hline
Street View only  & 0.42 & 0.40 & 0.32 \\ \hline
Aerial only      & 0.55 & 0.45 & 0.32 \\ \hline
Attrib.\ + Street   & 0.11 & 0.11 & 0.11 \\ \hline
Attrib.\ + Aerial   & 0.10 & 0.11 & 0.11 \\ \hline
Full model          & 0.10 & 0.09 & 0.08 \\ 
\end{tabular}}
\vspace{1em}
\\
\resizebox{.95\textwidth}{!}{%
\begin{tabular}{r|c|c|c}
    $R^2$           & 4-layers & 8-layers & 13-layers \\ \hline
Attributes only   & 68.96 & -- & -- \\ \hline
Street View only & 2.65 & 1.66 & 3.73 \\ \hline
Aerial only     & 6.24 & 5.12 & 5.00 \\ \hline
Attrib.\ + Street  & 71.37 & 68.78 & 72.70 \\ \hline
Attrib.\ + Aerial  & 71.61 & 73.18 & 75.86 \\ \hline
Full model         & 72.81 & 73.98 & 76.51 \\ 
\end{tabular}}
\end{minipage}
\end{table}

\subsection{Generalization: Holding out Southwark}

In the second experiment, we split the dataset so the entire borough of Southwark in Figure~\ref{fig:southwark} becomes a spatially out-of-sample test set. By splitting the dataset over the entire borough we show that the image network is not simply memorizing locations and recognizing neighboring streets as having similar house prices. This is a very difficult challenge, which tests the ability of the learned network to generalize to new locations which may have different visual cues indicating the desirability of neighborhoods.

Of particular note is the fact that the introduction of visual latent response do not just substantially improve the accuracy of the regressor, but also the stability when generalizing to unseen regions of London. Although all models exhibit a significant drop off when forced to generalize to a missing London borough rather than simply to data missing at random, this loss in accuracy is cut by two thirds --- only dropping by around 5\% rather than 15\% --- when using regressors that make use of attributes and visual latent response.   
This is remarkably successful given the challenge of the task and the high visual diversity of boroughs of London.

\subsection{Linear Hedonic Pricing Comparison}
\label{sec:linear-experiment}
In the third experiment, we compared the linear hedonic price model which is a linear combination  of both housing attributes $X$ and the image attributes $F(S)$, $G(A)$ with the traditional linear hedonic price regression model of using only housing attributes.

We fit the linear hedonic price regression model both with and without proxy variables for visual urban appearance, following equations \eqref{eq:hedonic1} and \eqref{eq:hedonic2}. 
The result, included in Table~\ref{tab:run4}, shows that the linear model with proxy variables offers a significant improvement over the standard model, coming much closer to the accuracy of the more general hedonic perceptron 
while retaining the interpretability of the linear model;
in fact, the linear model with visual desirability included performs similarly to the attributes-only hedonic perceptron.

\begin{table}[]
\centering
\caption{Hedonic Model Results\label{tab:run4}}

\begin{tabular}{r|c|c|c|c}
\textbf{}        & \multicolumn{2}{c|}{Random} & \multicolumn{2}{c}{Southwark} \\ \hline
\textbf{}        & $R^2$      & $MSE$     & $R^2$       & $MSE$       \\ \hline
Linear (Attrib.)     & 72.50\%          & 0.09              & 62.73\%            & 0.14               \\ \hline
Linear (Attrib.+Vis) & 76.93\%          & 0.08              & 67.85\%            & 0.12               \\ \hline
Additive (Attrib.) & 80.04\%            & 0.07              & 66.82\%            & 0.11               \\ \hline
Additive (Attrib.+Vis) & 83.54\%        & 0.06              & 72.68\%            & 0.09               \\ \hline
XG.Boost (Attrib.) & 81.72\%        & 0.06              & 67.78\%            & 0.11               \\ \hline
XG.Boost (Attrib.+Vis) & 84.13\%        & 0.05              & 74.23\%            & 0.09              \\ \hline
NonLin (Full model) & 84.67\%          & 0.05              & 76.51\%            & 0.08               \\ 
\end{tabular} \vspace{0.5em}
\end{table}

To demonstrate how interpretable our new approach is, we plot on a map the values 
$\gamma V(F, G)$
from the full model, for Street View and satellite data across the whole of central London, including areas for which we have no transaction data available. This map, shown in Figure~\ref{fig:residue2}, contains the predicted contribution to the hedonic utility of properties based on their visual appearance. 

\begin{figure}
	\centering 
	\includegraphics[width=0.9\linewidth] {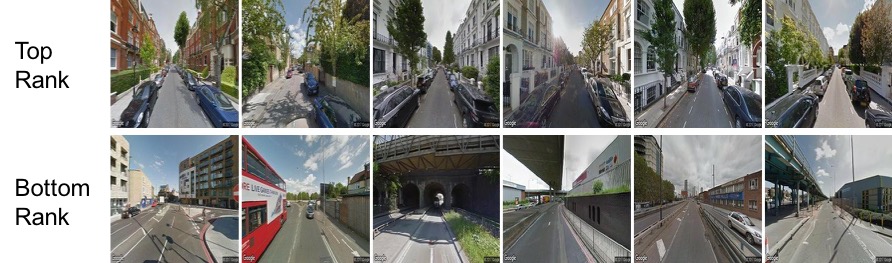}
	\caption{
	Top and bottom ranked images sorted by visual desirability score. Top-ranked images tend to feature narrower streets, with a variety of building frontages and plenty of vegetation. Low-ranked images feature wider motorways, overpasses, empty street frontage, and a lack of vegetation.}
	\label{fig:top_bottom_rank}
\end{figure}

To elaborate, we included the top and bottom ranked Street View images sorted by visual desirability score $\gamma V(F, G)$ in Figure \ref{fig:top_bottom_rank}. The top ranked images highlight higher density Georgian and Victorian buildings with large white-sashed windows, tree-lined streets and  brick cladded facade exterior. On the contrary, the bottom ranked images have generally smaller windows, wider streets at lower density, no greenery and blank frontages. 

\paragraph{Training the visual desirability network.}
A subtlety here lies in the training objective for the visual desirability network $V(F, G)$.
Informally, we would like the network which processes the image data $S$ and $A$ to learn a feature which only includes visual desirability, and not any properties which one might be able to predict from the images.
To reduce the possibility of confounding, we train the network in a two-stage process in order to learn a proxy variable which is orthogonal to the known housing attributes $X$.
First, we fit a nonlinear hedonic perceptron model $H(X)$ which predicts the log housing price from the attributes, but not including the images. We then compute residuals $R = Y - \hat Y$, where $\hat Y = H(X)$ is the predicted log price.
The network defined by $V(F(S), G(A))$, with weights $W_S, W_A$, and $W_V$, is then trained to predict these residuals $R$.
The original hedonic perceptron model $H(X)$ is then discarded, and the scalar output $V(F(S), G(A))$ which was used to predict the residual can then be used as an input into the linear model alongside the columns of $X$.
This can be interpreted as a method for fitting a generalized additive model in which price is assumed to be a sum of two overall components; one an unknown function of the housing attributes, and the other an unknown function of the images. The multi-step process by which the image feature network $V(F(S), G(A))$ is fit conditioned on the predictions of the hedonic perceptron $H(X)$, which is itself then re-estimated (albeit with an alternative interpretable model definition), is an instance of the more general back-fitting algorithm for additive models \citep{buja1989linear}. 

\subsection{Generalized Additive Model}

In the fourth experiment, we fit a generalized additive model (GAM), where the response variables $X$ are the sum of its smoothing functions, with and without the visual feature 
$V(F, G)$
\cite{Wood2006}. 
Among several basis functions, the model utilizes a set of thin plate regression splines as proposed in \cite{Wood2006}, which can avoid subjective knot placement, following the works of \cite{helbich2013}. The degree of smoothing is controlled by a smoothing parameter which is optimised by minimising the generalized cross-validation score (GCV). Its nonlinearity is measured by the effective degree of freedom (EDF) where 1 signals a linear relationship and >1 signal a nonlinear relationship \cite{Wood2006}. The GAM model effectively becomes a linear additive model when EDF is equal to 1.

We start from the linear model without splines where the results in Table~\ref{tab:run5} shows a significant improvement in goodness of fit as measured by Akaike information criterion $AIC$\footnote{AIC is a metric for model comparison which considers both the goodness of fit and the number of estimated parameters in the model. The metric is computed as follows $AIC=2k-2L$ where k is the number of estimated parameter and L is the log-likehood value of the model} when we include the visual desirability variable. All housing attribute is statistically significant at the $p<0.01$ level. For the linear additive model, a $0.1$ unit change in the visual desirability variable $[-0.21,0.30]$ is equal to approximately $13\%$ change in house price. The result from the GAM model with splines in Table~\ref{tab:run6} shows a significant improvement in $AIC$ when compared to the linear model. While the non-parametric effect from each housing attribute is also statistically significant at the $p<0.01$ level. To translate the GAM results, we can take the ratio between the natural exponentiated $max$ and the $min$ predictions of the GAM model for the visual latent response. We find that the differences in visual desirability throughout its entire range can add/reduce up to $1.85$ times of house price.

\begin{figure}
	\centering 
	\begin{minipage}[b][][b]{0.7\linewidth}
	\includegraphics[width=0.99\linewidth,trim={0 4in 0 0},clip]{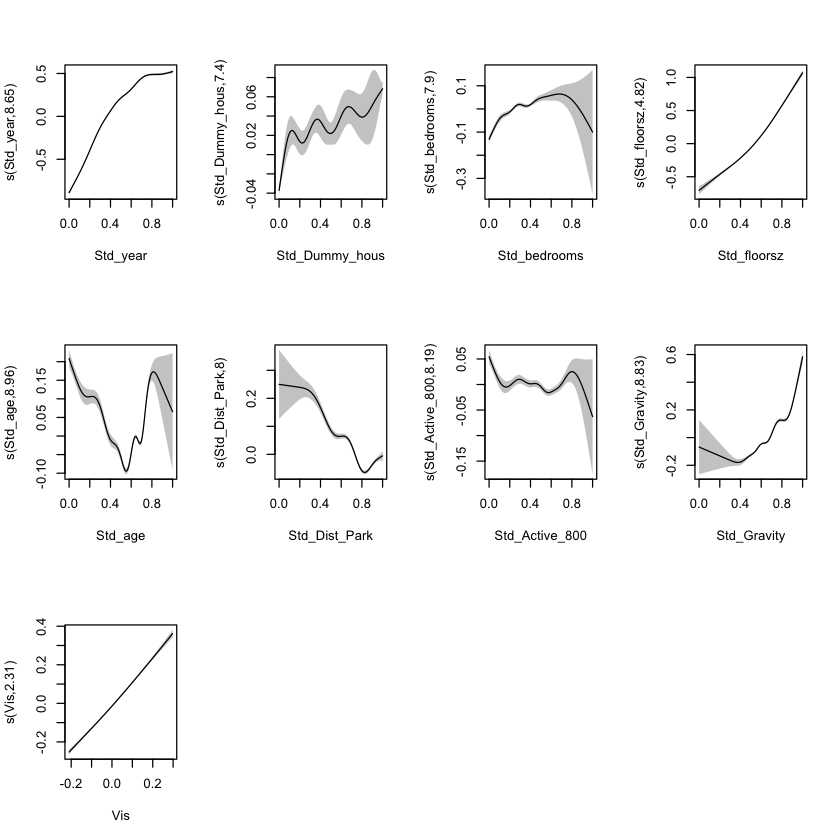}
	\end{minipage}\hfill
	\begin{minipage}[b][][b]{0.25\linewidth}
    \includegraphics[width=0.9\linewidth,trim={0 0 8in 8in},clip]{images/R-GAM06.png}
    \vspace{0.5in}
	\end{minipage}
	\caption{Partial dependence plot for the GAM. Housing attributes are shown on the left, and the visual feature summary on the right.}
    \label{fig:GAM_plot}
\end{figure}

Due to its additive nature, we can interpret individual features by holding other features at their mean. The plot, shown in Figure~\ref{fig:GAM_plot}, are the partial dependency plots of the $GAM$.\footnote{ 
Note that the partial dependencies for the housing attributes are unaffected by whether the visual latent response are included (and thus we do not include these as a separate plot). This is because the housing features are orthogonal to the visual desirability feature, due to the two-stage procedure used to train our model.}

The plot shows that the size of the property, year, accessibility, and housing type are (as expected) significantly positively related to house price, while the distance to a park is significantly negatively related to house price. As shown in the dependency plots, some of the housing features have more complex nonlinear relationships with house price. One of the most apparent is the nonlinear relationship with age, where the depreciation effect from physical deterioration is followed by an appreciation effect of older dwellings \cite{Brunauer2010}. The plot for bedroom counts shows the diminishing returns of having more and more extra bedrooms. On the other-hand, the plot for shopping amenities illustrates a more complex trade-off between both the positive and negative effects: the upward sloping curve before the last quartile shows the benefits attained from having very high-levels of shopping amenities can potentially outweighs its costs, while the downward sloping curve in the first quartile shows the costs attained from having very low-levels of shopping amenities can potentially outweights its benefits. Due to the lack of samples at the tail of the distribution, the confidence for bedrooms, shopping amenities and age diminishes.

\color{black}

\begin{table}[]
\centering
\caption{
Results of the linear model without splines. The result shows significant improvement in goodness of fit as measured by AIC when visual desirability is included into the model. All house attribute is statistically significant at the p<0.01 level. \label{tab:run5}}
\begin{tabular}{l|c|c|c|c|c|c}
\textbf{} & \multicolumn{3}{c}{\textit{Base.}} & \multicolumn{3}{c}{Vis.} \\ \hline
\textit{Par} & Est & Std.E. & T-val. & \textit{Est} & \textit{Std.E.} & \textit{T-val.} \\ \hline
Inter. & 10.016 & 0.024 & 416*** & 10.12 & 0.022 & 466*** \\
year & 1.618 & 0.008 & 208*** & 1.599 & 0.007 & 228*** \\
house & 0.184 & 0.005 & 35*** & 0.110 & 0.005 & 23*** \\
bed & -0.153 & 0.024 & -6*** & 0.082 & 0.022 & 4*** \\
size & 2.164 & 0.024 & 89*** & 1.856 & 0.022 & 83*** \\
age & 0.062 & 0.013 & 5*** & -0.08 & 0.012 & -7*** \\
park & -0.598 & 0.016 & -37*** & -0.437 & 0.015 & -30*** \\
shops & -0.121 & 0.012 & -10*** & -0.048 & 0.011 & -5*** \\
gravity & 1.038 & 0.020 & 53*** & 0.953 & 0.018 & 54*** \\
Vis. &  &  &  & 1.36 & 0.017 & 82*** \\ 
\hline
\hline
AIC & \textit{12353} &  &  & \textit{6144} & \textit{} & \textit{} \\
GCV & \textit{0.089} &  &  & \textit{0.072} &  & \\
\hline 
\hline \\[-1.8ex] 
\textit{Note:}  & \multicolumn{3}{r}{$^{*}$p$<$0.1; $^{**}$p$<$0.05; $^{***}$p$<$0.01} \\ 

\end{tabular}

\end{table}

\begin{table}[]
\centering
\caption{
Results of the Generalised Additive Model with splines. The model utilizes a set of thin plate regression splines where its smoothness is controlled by minimising the generalized cross-validation score (GCV). Its nonlinearity is measured by the effective degree of freedom (EDF) where 1 signals a linear relationship and greater than 1 signals a nonlinear relationship \cite{Wood2006}. The result shows significant improvement in goodness of fit as measured by $AIC$ and the non-parametric effect from each housing attribute is statistically significant at the $p<0.01$ level.  
\label{tab:run6}}
\begin{tabular}{l|c|c|c|c|c|c}
\textbf{} & \multicolumn{3}{c}{\textit{Base.}} & \multicolumn{3}{c}{Vis.} \\ \hline
\textit{Par} & Est & Std.E. & T-val. & \textit{Est} & \textit{Std.E.} & \textit{T-val.} \\ \hline
Inter. & 12.03 & 0.002 & 8022*** & 12.03 & 0.001 & 9069*** \\
 \multicolumn{1}{l}{}  & \multicolumn{1}{l}{} & \multicolumn{1}{l}{} & \multicolumn{1}{l}{} & \multicolumn{1}{l}{} & \multicolumn{1}{l}{} & \multicolumn{1}{l}{} \\ \hline
N.Par & EDF & R.df. & F-val. & EDF & R.df & F-val. \\ \hline
year & 8.49 & 8.93 & 6624*** & 8.66 & 8.97 & 7878*** \\
house & 7.78 & 8.55 & 159.4*** & 7.33 & 8.22 & 76.64*** \\
bed & 8.46 & 8.91 & 25.36*** & 7.84 & 8.63 & 38.02*** \\
size & 5.89 & 7.15 & 1259*** & 4.90 & 6.14 & 1362*** \\
age & 8.97 & 9.00 & 156.02*** & 8.96 & 9.00 & 160.85*** \\
park & 8.21 & 8.84 & 176.5*** & 8.03 & 8.76 & 118.90*** \\
shops & 7.00 & 8.06 & 26.28*** & 8.09 & 8.77 & 11.24*** \\
gravity & 8.82 & 8.99 & 315.1*** & 8.82 & 8.99 & 364.1*** \\
Vis. &  &  &  & 2.31 & 2.92 & 2258*** \\ 
\hline
\hline
AIC & \textit{3786} &  &  & \textit{-2252} & \textit{} & \textit{} \\
GCV & \textit{0.067} &  &  & \textit{0.054} &  & \\
\hline 
\hline \\[-1.8ex] 
\textit{Note:}  & \multicolumn{3}{r}{$^{*}$p$<$0.1; $^{**}$p$<$0.05; $^{***}$p$<$0.01} \\ 

\end{tabular}
\end{table}

In addition, we have included 
a Gradient Boosting ($XG.Boost$) regression as a baseline for the non-linear models to predict log house price from the housing attributes with and without the visual response. Gradient boosting regression is a state of the art regression technique which works by sequentially fitting the new predictors to the residuals of the previous predictors. Specifically, we use the $XGBoost$ library for the gradient boost regression with default parameters \cite{XGBoost}. The extended results in Table~\ref{tab:run4} shows the mean squared error and $R^2$ comparing the linear model, the additive model, 
the gradient boosting model and the full nonlinear model. The result shows all the models achieve better accuracy with the visual latent response than without. The results are also significantly better for the Southwark held-out experiments suggesting the visual response potentially helps with generalisation in unseen context. More importantly, the additive model which has greater interpretability achieves comparable accuracy when compare to the other two nonlinear models. 


\section{Discussion and Conclusion}

We have presented a novel approach to house pricing that leverages visual knowledge of the urban environment to improve predictive power. In contrast to previous work \cite{Peterson2009,Ahmed2016,Poursaeed2018} that have made use of images of the interior and exterior of the property for sale, we have focused on characterizing the local street neighbourhood where the property sits on, and with the property making up only a small proportion of the aerial images; while the Street View images we make use of typically do not contain the property itself.

This study finds encouraging results in predicting house prices in London using street images both at ground level and at aerial level. We find that the traditional housing attributes explains the majority of the variance of house price; we also find that the model augmented with features extracted from images performs better than the model without image features. Augmenting the baseline housing attribute model with aerial images perform better than the baseline model with ground-level Street View photos, suggesting that buyers might be valuing a visually desirable neighborhood more than a visually desirable street. Importantly, we have developed a visual proxy measure that improves explainability with only minor losses in accuracy.

There is much promising future work and further directions for exploration. For example, the focus on London raises questions on whether the proxy for visual desirability generalizes to different markets. 
Comparison between cities by fitting this model to other data could potentially reveal differences; allowing, for example, quantitative exploration of ways in which aesthetic preferences differ between London and Kyoto. 

Additional research is also needed to visually explain the convolutional neural network model. For example, extracting discriminative features between higher and lower house price from street images can potentially bring greater clarity to the model \cite{Arietta2014} or the use of  visual explanation methods such as Grad-CAM \cite{GradCAM2016}, LIME \cite{Ribiero2016} and Deeplift \cite{Shrikumar2017} to identify the regions of the image that gets activated for higher or lower visual desirability score.

Beyond this, the images from Google Street View and Bing Aerial photos are not entirely reliable. Concerns can range from visual obstruction, poor lighting condition and differences in weather can affect the result. 
However, we believe that these issues are distributed fairly uniformly through the dataset and uncorrelated with price or housing attributes; so while they may introduce noise, we do not expect them to introduce any systemic bias.

The work could be extended by making use of additional complementary cues, such as the images of the property interior \cite{Ahmed2016,Poursaeed2018}, the views from within the property \cite{Seresinhe2017}, and the text description of the property. We can test and fuse these multi-modal features using different learning methodologies.
Another notable opportunity concerns confounding environmental variables not accounted for in the hedonic price model. Additional environmental cues such as urban density, green foliage \cite{Li2018}, and pollution levels\cite{McCord2018B} should be incorporated into a future model. 
We would also hope that the methodology used here, to learn a visual latent response from unlabeled and unstructured data, could be adapted to learn other important, interpretable cues from additional raw data sources.

Future research will also aim to capture wider neighbourhood and district level visual desirability effect. This research could connect with the econometric literature on capturing second order spatial autocorrelation effects \cite{Anselin1988} using locally weighted regression methods \cite{McCluskey2013} or more recently in using graph embedding methods in capturing similar neighbourhood effect \cite{You2017}.

As our aim of the research is (a) to retrieve an interpretable visual desirability score and (b) to compare a non-interpretable and a semi-interpretable model with and without visual latent response, we consider the multi-layer-perceptron as a suitable baseline for our research. However, there is a clear need to incorporate a more diverse sets of methods and feature cues into the research. A key limitation for comparative research in the real estate is the geographical specificity of the model and the lack of a common baseline dataset to test different methods against. As a result, future research is needed within the community to develop a consistent, open, multi-modal and geographically diverse database.

Developing more reliable house price models is an important topic for urban planning and the real estate sector.  The implication is that these models can be used to improve existing mass appraisal models \cite{McCluskey2013} which can be used in mortgage assessment. Secondly, these models can also be used to improve the visual desirability of existing streets and new neighborhoods through housing policy and planning guidance.
\color{black}

Our use of end-to-end training has allowed us to avoid the need for costly annotation of urban data, while still extracting meaningful proxy values from the urban environment. As well as improving the accuracy of standard models we believe that these visual proxies will be of interest to economists on estimating the willingness to pay for different levels of visual desirability. To that end we are both releasing the training code, allowing these features to be developed in new environments, and the pre-trained models, allowing the automatic generation of such proxy values in urban environments similar to London.

\begin{acks}
This work was supported by The Alan Turing Institute under the UK Engineering and Physical Sciences Research Council (EPSRC) grant no. EP/N510129/1. The authors would also like to thank the anonymous referees for their valuable comments. 
  
\end{acks}

\bibliographystyle{ACM-Reference-Format}
\bibliography{KDD2018_Law_et_al} 

\end{document}